# User-Level Memory Scheduler for Optimizing Application Performance in NUMA-Based Multicore Systems

Geunsik Lim
*Software Center*
*Samsung Electronics*
*Suwon 443-370, Republic of Korea*
geunsik.lim@samsung.com

Sang-Bum Suh
*Software Center*
*Samsung Electronics*
*Suwon 443-370, Republic of Korea*
sbuk.suh@samsung.com

*Abstract*—**Multicore CPU architectures have been established as a structure for general-purpose systems for high-performance processing of applications. Recent multicore CPU has evolved as a system architecture based on non-uniform memory architecture. For the technique of using the kernel space that shifts the tasks to the ideal memory node, the characteristics of the applications of the user-space cannot be considered. Therefore, kernel level approaches cannot execute memory scheduling to recognize the importance of user applications. Moreover, users need to run applications after sufficiently understanding the multicore CPU based on non-uniform memory architecture to ensure the high performance of the user's applications. This paper presents a user-space memory scheduler that allocates the ideal memory node for tasks by monitoring the characteristics of non-uniform memory architecture. From our experiment, the proposed system improved the performance of the application by up to 25% compared to the existing system.**

## I. INTRODUCTION

As users require a high performance computer [1] for mass computation, applications need a high-performance computing system to execute jobs more quickly. The demand for high performance has meant CPU architecture has developed from a single CPU to a multicore CPU. Recently, CPU architecture has evolved to a multicore CPU architecture based on non-uniform memory architecture as shown in Figure 1. In response to this, the design of the modern computer faces a very challenging software assignment called thread scheduling. The technique is used to control the memory nodes at the operating system level, and manages the memory usage of tasks to avoid the problem where tasks lean towards one memory node.

However, in the technique of using the operating system level the relative importance among user applications cannot be recognized. Therefore, it is important that we find an automatic memory scheduling method in the user-space. Moreover, users need to have in-depth system knowledge to obtain high performance applications and effective memory utilization in the existing systems. Therefore, users cannot utilize the multicore system based on non-uniform memory architecture (NUMA) [2], [3] because they need to have an in-depth NUMA architecture knowledge. This paper presents a novel memory scheduler that removes unnecessary memory latency and supports high-performance execution of the application. Our proposed system schedules memory nodes after monitoring NUMA architecture automatically in the user-space.

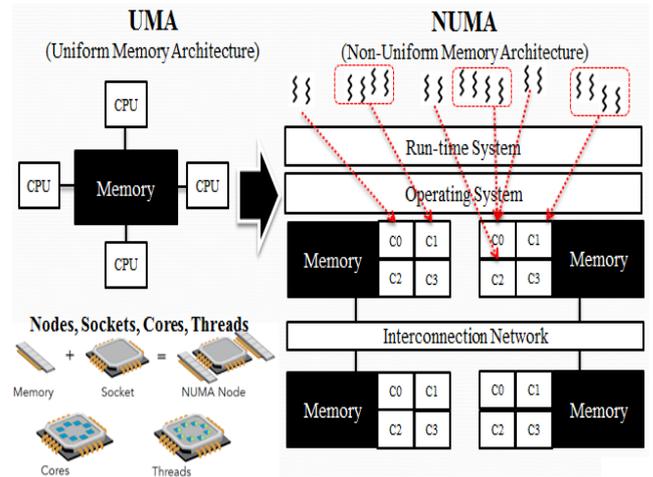

**Figure 1. Evolution of memory architecture**

The remainder of this paper is organized as follows. Related work is described in Section II. Section III addresses the design and implementation of the proposed techniques in detail. Section IV shows the evaluation results. Finally, Section V concludes the paper.

## II. RELATED WORK

### A. NUMA scheduling techniques in kernel-space

*SchedNUMA* [4] optimizes memory locality in the NUMA system by placing tasks [5] into the same NUMA node. However, this approach cannot maintain the compatibility of API because the approach needs to perform definition and implementation of additional APIs [6] for grouping processes. *Automatic NUMA Balancing* [7] migrates the pages of a task to another memory node when the page fault handler attempts to swap the unmapped pages from the page table entry. This technique can execute memory scheduling automatically in the kernel space. However, it is very difficult for the operating system to optimize the optimal NUMA tuning.. Ultimately, the system administrator needs to fulfill tuning work in addition to the optimization tools.

## B. NUMA scheduling techniques in user-space

*Sergey Blagodurov* [8] addressed the optimization method of performance that uses CPU affinity functions [9] in the user-space to obtain maximization of the application's performance. However, this technique damages the effective memory utilization of tasks because the proposed idea statically fixes tasks into a specific NUMA node. Moreover, this approach does not handle the automated method of the memory scheduling in the user-space.

## III. PROPOSAL OF USER-LEVEL NUMA-AWARE MEMORY SCHEDULER

Our proposed technique maintains an ideal memory locality to help the high-performance execution of the application by removing the possibility of memory latency. To reach this goal, the proposed system automatically executes (re)allocation of jobs by finding the best ideal NUMA node in the user-space with the collected information after monitoring the NUMA bus topology and run-time memory usage.

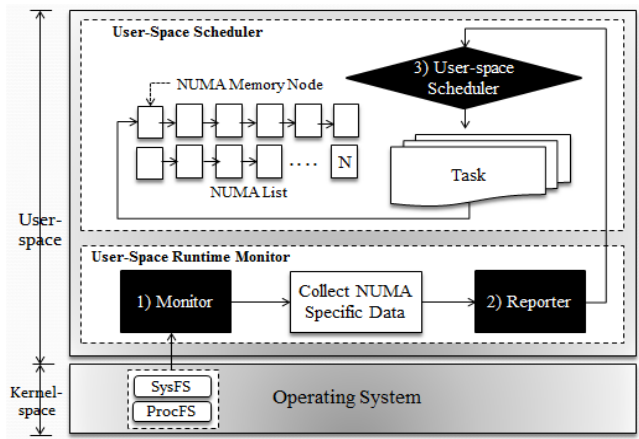

**Figure 2. System Architecture of our proposed system**

Figure 2 shows the system architecture that remaps tasks into an ideal memory range after monitoring the characteristics of the NUMA architecture [10] in the user-space. The proposed system consists of three major components as follows:

1) Runtime monitor to monitor memory usage based on NUMA topology.
2) Reporter to submit NUMA related data to the scheduler of the user-space using the collected information.
3) User-space memory scheduler to (re)allocate tasks to the ideal memory node

We implemented the proposed idea with three algorithms. *Algorithm 1* of figure 3 collects the scheduling related data by scanning the proc file system (*procfs*) and sys file system (*sysfs*) [11]. *Algorithm 2* of figure 4 reports only NUMA specific data to the user-space NUMA scheduler after filtering the collected schedule data. Finally, *algorithm 3* of figure 5 moves a task to the ideal memory node with the collected information.

## A. Algorithm of the runtime monitor

The proposed system creates a thread to collect scheduling information of the tasks from the file system such as *procfs* and *sysfs*. The created thread repeatedly executes work collecting scheduling information until in 3) the user-space NUMA scheduler is completed.

**Algorithm 1. Monitor: Runtime monitoring mechanism**
1. Create a new thread for receiving and dealing with the run-time monitoring data
2. **Repeat monitoring** until user-space NUMA scheduler stops
3.     Sleep for an NUMA specific data
4.     Collect the data monitored from *proc* file system (/proc/<pid>/{stat | numa_maps})
5. **End Repeat loop**

**Figure 3. Algorithm of the runtime monitor**

## B. Algorithm of the reporter

In 2), the Reporter receives monitoring information of the runtime until in 1) the Monitor terminates. First, in 2), the Reporter preserves information that filters NUMA specific data from the collected data. If the distribution status of the memory node is not balanced or the execution flow of tasks is changed by the operating system level memory scheduler, the Reporter in 2) executes the assignment to find a suitable memory node for the high-performance of important applications. Second, in 2), the Reporter calculates runtime high-performance factors, re-sorting processes of the NUMA list, and the ideal memory node for new tasks. Finally, in 2), the Reporter sends this information to 3), the user-space NUMA scheduler.

**Algorithm 2. Reporter: Mechanism to report collected NUMA specified data**
*Input: run-time monitoring data*
1. **Repeat until** runtime monitoring mechanism stop
2.     Receiving data and filtering them from online monitoring
3.     Collect NUMA specific data
4.     **If** loading of system is unbalanced or behavior of the processes changed or powerful core is idle
5.         Computing the Run-time speedup factor
6.         Sorting the process NUMA list by multi-core speedup factor
7.         Computing the contention degradation factor
8.         Sorting the process NUMA list by contention degradation factor
9.         Sending signal to trigger schedule
10.     **End if**
11. **End Repeat loop**

**Figure 4. Algorithm of the reporter**

## C. Algorithm of the user-space memory scheduler

**Algorithm 3. User-space scheduler: Automatic NUMA aware scheduling**
*Input: NUMA list*
1. Computing the number of powerful core candidate based on **load balanced memory policy**
2. Retrieving suitable processes to be scheduled on powerful cores from NUMA list
3. Setting static CPU pin from manual input of administrator
4. **If** retrieved processes != current processes on powerful cores
5.     Migrate the processes
6. **End if**
7. **If** current resource contention degradation is too big
8.     Scatter the processes with heavy contention
9.     Calculating degradation factor in order to minimize resource contention degradation
10.     Migrate the processes and the its sticky pages
11. **End if**

**Figure 5. Algorithm of the user-space scheduler**

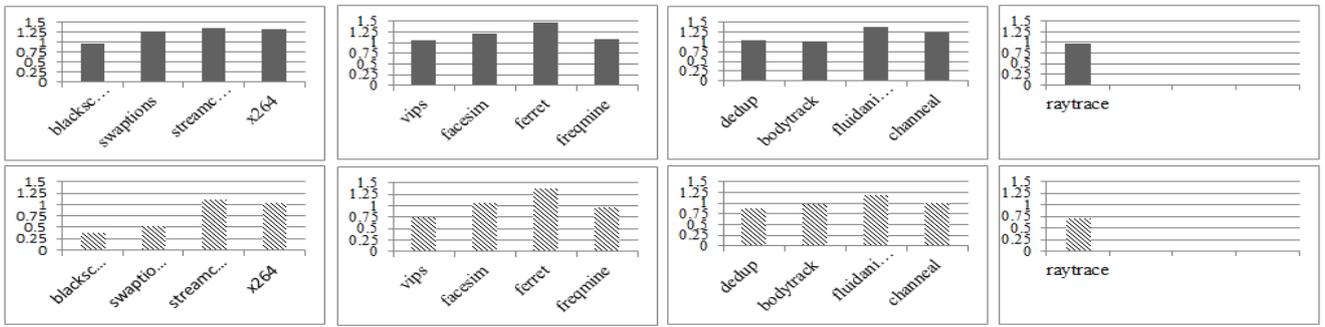

**Figure 6. Accuracy of the performance degradation factor (upper figures refer to performance degradation due to contention; lower figures refer to contention degradation factor).**

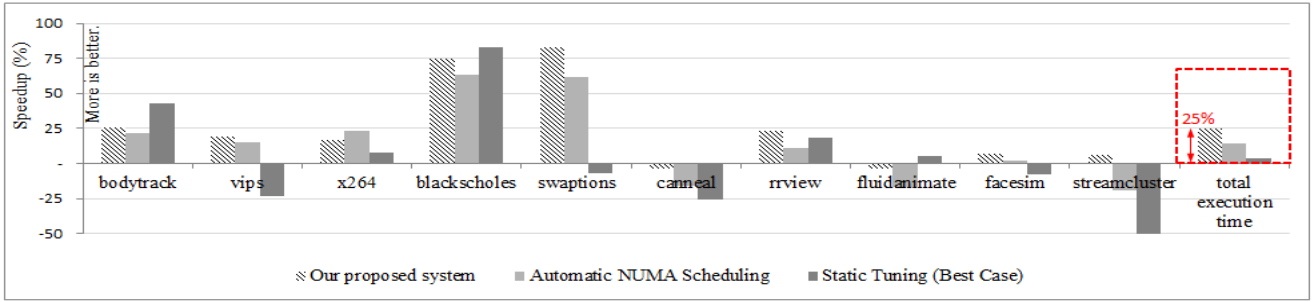

**Figure 7. Speedup of our proposed system, the Automatic NUMA Scheduling, and the Static Tuning on a 40-cores platform.**

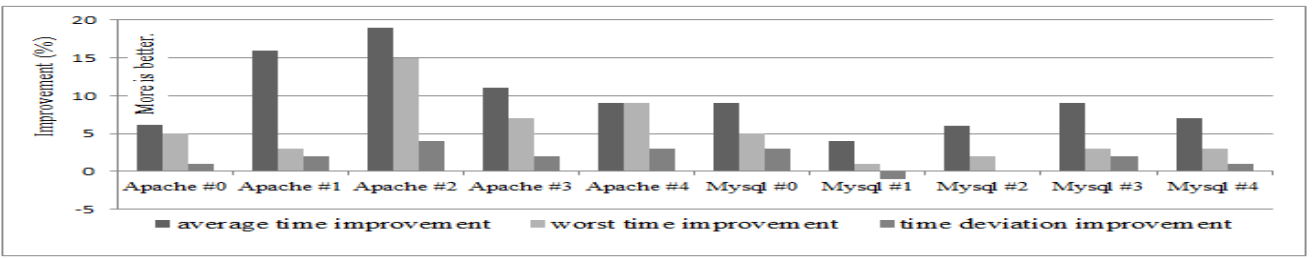

**Figure 8. Performance experiment result in NUMA system using Apache webserver and MySQL database.**

In 3), the User-space scheduler determines a suitable CPU number to ensure high-performance of a specified user-space application using information received from 2), the Reporter. At this time, in 3), the User-space scheduler considers the static CPU affinity information required by the server administrator as well as the NUMA specific information received by 2), the Reporter. Moreover, if the contention of the resource is too high, in 3), the user-space scheduler attempts to distribute the tasks, considering the contention ratio of the task.

## IV. EVALUATION

We prepared an experimental system with an Intel Xeon NUMA Server (*Server: DELL PowerEdge R910, CPU: Intel Xeon E7-4850 @2.00GHz 40 Cores, Memory: 32GiB, OS: Linux 3.2, Platform: Ubuntu 12.04 LTS 64bit*) to verify the effects of our proposed idea. We set an evaluation system as follows to provide a reasonable experimental condition:

• Half of the workload focuses on the CPU intensive task scheduling with the PARSEC benchmark suite [3], [12].

• The other half of the workload focuses on memory-intensive task scheduling with the PARSEC benchmark suite.

**Table 1. Key characteristic of PARSEC benchmarks**

| Program | Application domain | Parallelization model | Parallelization granularity | Data sharing | Data exchange |
|---|---|---|---|---|---|
| blackscholes | Financial analysis | data-parallel | coarse | low | low |
| bodytrack | Computer vision | data-parallel | medium | high | medium |
| canneal | Engineering | unstructured | fine | high | high |
| dedup | Enterprise storage | pipeline | medium | high | high |
| facesim | Animation | data-parallel | coarse | low | medium |
| ferret | Similarity search | pipeline | medium | high | high |
| fluidanimate | Animation | data-parallel | fine | low | medium |
| freqmine | Data mining | data-parallel | medium | high | medium |
| streamcluster | Data mining | data-parallel | medium | low | medium |
| swaptions | Financial analysis | data-parallel | coarse | low | low |
| vips | Media processing | data-parallel | coarse | low | medium |
| x264 | Media processing | pipeline | coarse | high | high |

PARSEC [12] is a benchmark suite composed of multithreaded programs. The suite focuses on emerging workloads [13] and was designed to be representative of next-generation shared-memory programs for chip-multiprocessors. Table 1 shows a qualitative summary of the inherent key characteristics of PARSEC benchmarks. The pipeline model is a data-parallel model which also uses functional partitioning. PARSEC workloads were chosen to cover different application domains, parallel models [14], and runtime behaviors.

Figure 6 shows the correlation of performance reduction between the imbalance of memory utilization and memory contention in the NUMA architecture based multicore system. When we applied our approach to the NUMA system, PARSEC incurred a performance reduction of over 90%. This means that PARSEC is very suitable as a workload generating memory contention in a NUMA architecture based multicore system [15].

Figure 7 shows the experimental result of the application execution time of our proposed system compared to the existing system. From our experiment, our proposed system improved the execution time of PARSEC applications by up to 25% when we adapted our proposed system in the existing system. The proposed system could obtain 85% of improved execution time compared to Automatic NUMA Scheduling because our proposed system can recognize the importance of user-space applications. The Static Tuning technique manually optimizes tasks [16] with the CPU affinity technique, and had good results at the three applications, including *bodytrack*, *blackscholes*, and *fluidanimate* [12]. This means that the Static Tuning technique is not practical and depends on the technical ability of the server administrator. Therefore, we were not able to obtain consistent results for performance improvement with the Static Tuning method.

Additionally, we evaluated the throughput of the webserver and database in the real server environment that executes many service daemons to verify effects of our proposed system. Figure 8 shows the experimental result of performance when we ran the Apache webserver and MySQL database. The y-axis of figure 8 refers to the average time for improvement, the worst time for improvement, and the time deviation for improvement. From our experiment, we verified that our proposed system could improve the throughput by up to 12.6% of the Apache webserver and 7% of the MySQL database without any manual optimization.

The proposed system can improve the execution speed of a specified application dramatically in a server environment that runs memory-intensive applications because the proposed system can execute the memory scheduling of tasks while considering the importance of user applications. We verified that the proposed system is very practical and useful in NUMA architecture based high-performance computing servers.

## V. CONCLUSION

The NUMA architecture for a scalable multicore system [7], [18] facilitates the fast high-performance of user-space applications. However, the system administrator needs to perform tuning work with optimization tools because optimal NUMA tuning using the operating system is very difficult. It is impractical for the administrator to understand the memory architecture of the NUMA architecture to obtain stable memory utilization and high-performance. The proposed system reallocates tasks into an ideal memory node with collected information after monitoring the characteristics of the NUMA topology [19] in the user-space without a kernel space. Our approach does not depend on the operating system because the proposed idea rearranges the tasks in the user-space. In other words, our proposed system is a new user-level NUMA aware memory scheduler considering the memory utilization and optimization of performance without the processor affinity technique that damages the memory utilization.

REFERENCES

[1] D. J. DeWitt and J. Gray, "Parallel database systems: the future of high performance database systems," In CACM, 1992.
[2] Z. Majo and T. R. Gross, "Memory system performance in a NUMA multicore multiprocessor," In SYSTOR, 2011.
[3] C. McCurdy and J. S. Vetter, "Memphis: Finding and fixing NUMA-related performance problems on multicore platforms," In ISPASS, 2010.
[4] Linux Kernel Mailing List, "Sched NUMA Rewrite," https://lkml.org/lkml/2012/5/9/312/, May 2012.
[5] C. Banino, O. Beaumont, L. Carter, J. Ferrante, A. Legrand, and Y. Robert, "Scheduling strategies for master-slave tasking on heterogeneous processor platforms," IEEE Transactions on Parallel and Distributed Systems, pages 319-330, Apr. 2004.
[6] Linux Kernel Mailing List, "Scalability of signal delivery for POSIX threads," http://lkml.org/lkml/2004/11/22/432, Nov. 2004.
[7] Linux Kernel Mailing List, "Automatic NUMA Balancing v9," https://lkml.org/lkml/2013/10/7/136, Oct. 2013.
[8] S. Blagodurov, S. Zhuravlev, and M. Dashti, "A case for NUMA-aware contention management on multicore systems," In USENIX ATC, Jun. 2011.
[9] J. Rao, K. Wang, X. Zhou, and C. Z. Xu, "Optimizing virtual machine scheduling in NUMA multicore systems," In High Performance Computer Architecture (HPCA2013), pages 306-317, Feb. 2013.
[10] K. Keeton, D. A. Patterson, Y. Q. He, R. C. Raphael, and W. E. Baker, "Performance characterization of a quad Pentium Pro SMP using OLTP workloads," In ISCA, 1998.
[11] P. Mochel, "The sysfs Filesystem," In Linux Symposium, pages 313-326, Jul. 2005.
[12] C. Bienia, S. Kumar, J. P. Singh, and K. Li, "The PARSEC benchmark suite: Characterization and architectural implications," In ACM PACT, pages 72-81, 2008.
[13] S. Balakrishnan, R. Rajwar, M. Upton, and K. Lai, "The impact of performance asymmetry in emerging multicore architectures," In Annual International Symposium on Computer Architecture, pages 506-517, Jun. 2005.
[14] J. B. Andrews and C. D. Polychronopoulos, "An analytical approach to performance/cost modeling of parallel computers," Journal of Parallel and Distributed Computing, pages 343-356, Aug. 1991.
[15] R. Kumar, D. M. Tullsen, P. Ranganathan, N. P. Jouppi, and K. I. Farkas, "Single-ISA heterogeneous multi-core architectures for multithreaded workload performance," In Annual International Symposium on Computer Architecture, pages 64-75, Jun. 2004.
[16] J. Rao, K. Wang, X. Zhou, and C. Xu, "Optimizing virtual machine scheduling in NUMA multicore systems," In HPCA, Feb. 2013.
[17] A. Baumann, P. Barham, P.-E. Dagand, T. Harris, R. Isaacs, S. Peter, and T. Roscoe, A. Schüpbach, and A. Singhania, "The multikernel: a new OS architecture for scalable multicore systems," In SOSP, 2009.
[18] M. A. Bender and M. O. Rabin, "Scheduling Cilk multithreaded parallel programs on processors of different speeds," In ACM Symposium on Parallel Algorithms and Architectures, pages 13-21, Jul. 2000.
[19] R. P. LaRowe, Jr., C. S. Ellis, and L. S. Kaplan, "The robustness of NUMA memory management," In ACM Symposium on Operating System Principles, pages 137-151, Oct. 1991.